\documentclass[fleqn,usenatbib,useAMS]{mnras}

\usepackage{newtxtext,newtxmath}

\usepackage[T1]{fontenc}

\DeclareRobustCommand{\VAN}[3]{#2}
\let\VANthebibliography\thebibliography
\def\thebibliography{\DeclareRobustCommand{\VAN}[3]{##3}\VANthebibliography}

\usepackage{graphicx}	
\usepackage{amsmath}	

\newcommand{\msun}{{\,\rm M_\odot}}

\newcommand{\kms}{\,{\rm km}\,{\rm s}^{-1}}

\newcommand{\stt}{$\sin^{2}(2\theta)\,$}
\def\gsim{ \lower .75ex \hbox{$\sim$} \llap{\raise .27ex \hbox{$>$}} }
\def\lsim{ \lower .75ex \hbox{$\sim$} \llap{\raise .27ex \hbox{$<$}} }
\defcitealias{Venumadhav16}{VE16}
\defcitealias{Ghiglieri15}{GL15}

\title[{\it XRISM} and Sterile Neutrinos]{Anticipating the {\it XRISM} search for the decay of resonantly produced sterile neutrino dark matter}

\author[M. R. Lovell]{
Mark R. Lovell$^{1}$\thanks{E-mail: lovell@hi.is}
\\
$^{1}$Centre for Astrophysics and Cosmology, Science Institute, University of Iceland, Dunhaga 5, 107 Reykjav\'ik, Iceland}

\date{Accepted 2023 July 18. Received 2023 June 19; in original form 2023 April 4}

\pubyear{2023}

\begin{document}
\label{firstpage}
\pagerange{\pageref{firstpage}--\pageref{lastpage}}
\maketitle

\begin{abstract}
The sterile neutrino ($N_1$) features in multiple extensions of the standard model and is a compelling dark matter candidate, especially as the decay of $N_1$ with mass $m_\rmn{s}=7.1$~keV is a possible source for the unexplained 3.55~keV X-ray line reported in galaxy clusters. This particle will be accessible to the {\it XRISM} X-ray mission over the next 12 months. We revisit the physics behind $N_1$ and the uncertainty in its parameters. We compare predictions for the $m_\rmn{s}=7.1$~keV $N_1$ mixing angle, \stt, and half-mode mass, $M_\rmn{hm}$, as described in the $\nu$MSM standard model extension to existing X-ray observations and structure formation constraints. The strongest available constraints rule out $N_1$ as a dark matter candidate, and a more optimistic reading of the data prefers \stt$=5\times10^{-11}$ and $M_\rmn{hm}=3.5\times10^{8}$~$\msun$. We highlight that the most promising upcoming opportunity for a detection is to find a line of velocity dispersion $\sim500$~$\kms$ in the Virgo cluster with {\it XRISM}, and then draw up a list of future objects of study to determine: (i) whether the line is from dark matter generally, and (ii) if from dark matter, whether that candidate is indeed $N_1$.     
\end{abstract}




\section{Introduction}

The nature of dark matter remains one of the most important outstanding puzzles in astrophysics, cosmology, and particle physics. Constraints on candidate dark matter particles come in two forms: (i) astrophysical-cosmological constraints, which constitute the impact that a dark matter particle has on its environment, and (ii) particle constraints, which are laboratory experiments -- including collider generation and relic dark matter scattering off standard model particles -- and attempt to detect radiation from dark matter that decays or annihilates. The second set of constraints is a necessary condition for any claim to have identified the dark matter, and the first set is an important extra condition that applies to the subclass of models for which dark matter has self-interactions or a cutoff at mass scales $>10^{7}$~$\msun$. 

One such model of interest is the resonantly produced sterile neutrino \citep{Shi99,Laine08}. This particle has a mass at the keV scale, and constitutes part of a broader standard model extension that also has the potential to explain baryogenesis and neutrino oscillations \citep{Asaka05,Boyarsky09a}. It undergoes free-streaming to a degree that affects the number of dwarf galaxies \citep{Abazajian01a,Bode01,Lovell14} and generates an X-ray decay signal \citep{Shrock74,Pal82,Abazajian01a,Abazajian:01b}, thus it is subject to both the astrophysical constraints and the particle constraints discussed above. A candidate X-ray decay signal from a putative 7.1~keV-mass particle has been reported in galaxy clusters, the M31 galaxy and in the halo of the Milky Way \citep[MW, e.g.][]{Boyarsky14a,Bulbul14,Cappelluti17,Hofmann19}. The interpretation of these results as dark matter decay is controversial \citep[e.g.][]{Anderson14,Jeltema16,Dessert20}. Further light is expected to be shed on this subject by the launch of the JAXA {\it XRISM} mission in mid-to-late 2023 \citep{XRISM21}, where a detection of the line, and subsequently a measurement of the lines' velocity dispersion, would constitute a particularly compelling piece of evidence for a dark matter decay origin \citep{Lovell19c}.

If a detection of this line were made, the next step would be to determine whether this general dark matter decay signal corresponds specifically to a resonantly produced sterile neutrino or instead to a still more exotic dark matter candidate. This analysis constitutes computing the expected particle physics parameters consistent with the notional {\it XRISM} detection -- sterile neutrino mass and mixing angle -- and comparing these results to constraints from X-ray non-detections and from structure formation constraints. The resonantly produced sterile neutrino model consistent with the reported X-ray detections is especially suitable for this analysis, because relaxing X-ray constraints requires a smaller mixing angle, which in turn generates a large-scale cutoff in the linear matter power spectrum and thus stronger structure formation constraints \citep{Abazajian14,Lovell16}. 

In this paper we revisit the physics of resonantly produced sterile neutrino dark matter in the cosmological context. We discuss the relationship between the sterile neutrino parameters on the one hand and its decay/free-streaming properties on the other, while illustrating in particular the systematic uncertainty in the calculation of the free-streaming scale. We compare the decay rate and free-streaming results to existing X-ray and structure formation constraints. We highlight the putative sterile neutrino parameters that we might expect for a line detected by {\it XRISM}, and end by outlining the following steps required to determine whether resonantly produced sterile neutrinos are indeed the dark matter.

This paper is organized as follows. In Section~\ref{sec:rpsndm} we discuss the physics of sterile neutrinos as dark matter, and in Section~\ref{sec:res} we compute matter power spectra and present the possible model physics parameters. We compare the results to observations in Section~\ref{sec:obs} and draw conclusions in Section~\ref{sec:conc}. 

\section{Resonantly produced sterile neutrino dark matter}
\label{sec:rpsndm}

 In this section we discuss the status of sterile neutrinos generated through resonant production, which in principle is allowed in multiple extensions of the standard model. We focus on the particular case of the neutrino minimal standard model ($\nu$MSM), and make reference to alternative extensions where appropriate.

\subsection{Introducing three extra neutrinos}

The original motivation for sterile neutrinos comes from observations of two unexplained phenomena in neutrino physics: the absence of right-handed neutrinos and neutrino oscillations. First, neutrinos are the only standard model fermion that have only ever been observed to have left-handed chirality, therefore it is reasonable to posit the existence a counterpart right-handed chirality neutrino to match the rest of the standard model. This particle is known as a `sterile' neutrino because its right-handed chirality suppresses its coupling to the weak force, as compared to the left-handed `active' neutrino of the standard model, at least in the case where neutrinos are Dirac particles. The picture becomes more complex in the presence of Majorana masses; see \citet{Adhikari17} for a discussion. Such a neutrino would then be expected to explain the observation of neutrino flavour oscillations. These oscillations can only occur if the neutrinos have mass, whereas the standard model of particle physics assumes that neutrinos are strictly massless. Sterile neutrinos may then endow standard model neutrinos with mass through processes such as the seesaw mechanism. Under this condition the masses of the sterile neutrinos are unconstrained, and models have been proposed in which the mass could be as large as a TeV \citep{Humbert15} or as small as an eV, the latter of which is known as a `light sterile neutrino' and has been proposed to explain several neutrino experiment anomalies \citep{Athanassopoulos96,Mention11}; see also \citep{deGouvea05} for a more phenomenological discussion of mass generation.

One specific model of note is the $\nu$MSM, which adds exactly three sterile neutrinos to the standard model \citep{Asaka05,Boyarsky09a}. Under this model, two of the sterile neutrinos are responsible for facilitating neutrino oscillations. These two particles have short lifetimes and are almost degenerate in mass, and are much more massive that the third sterile neutrino, in a manner that reflects the mass differences of the active neutrino mass eigenstates; see \citet{Abazajian17} for a review of alternative mass generation mechanisms. 

This model is especially compelling when the model parameters are set such that they can explain two additional outstanding problems in particle physics: baryogenesis and dark matter. Baryogenesis is achieved through low-scale leptogenesis, in which the two more massive sterile neutrinos preferentially decay into antileptons that are subsequently converted into baryons via the weak interaction sphaleron process, and thus leads to the required excess of baryons over antibaryons. Moreover, the ability to generate an imbalance between leptons and antileptons -- known as the lepton asymmetry -- enhances the likelihood of active neutrinos oscillating into the lower mass sterile neutrino in a manner similar to the Mikheyev–Smirnov-Wolfenstein effect \citep{Wolfenstein:1977ue,Mikheev:1986gs} that alters the oscillation rates of active neutrinos; this process is known as {\it resonant-production} \citep{Shi99,Abazajian01a}. 

This sterile neutrino has a lifetime much longer than the age of the Universe, and via the oscillation method can be generated in sufficient quantities to match the measured dark matter abundance. It also decouples immediately from the primordial plasma on production -- a process known as freeze-in -- and therefore evades bounds from Big Bang nucleosynthesis (BBN) and the cosmic microwave background (CMB) that otherwise constrain light dark matter particles \citep{Sabti20,An22}. Finally, these three phenomena -- neutrino oscillations, baryogenesis, and dark matter -- can be explained simultaneously if the masses of the two more massive sterile neutrinos fall in the $\sim$GeV regime and the lower mass sterile neutrino is a million times less massive, and thus in the $\sim$keV regime. Throughout the rest of this paper we will refer to the GeV-scale heavier sterile neutrinos in this specific model as $N_2$ and $N_3$, and to the keV-scale sterile neutrino as $N_1$. 

\subsection{Production of sterile neutrino dark matter}
 
The effect of the lepton asymmetry on the production has an important impact on the permitted sterile neutrino parameters. In the absence of a lepton asymmetry, the production rate is set by the particle mass and the mixing angle from sterile to active neutrinos, \stt \citep{Dodelson94}. The resonance effect promotes the conversion of lower energy active neutrinos to sterile neutrinos, thus lowering the velocity dispersion of the sterile neutrinos compared to what would be expected if the resonance were absent. This lepton asymmetry can be parametrized in multiple different ways, such as the difference in lepton and antilepton densities prior to dark matter production divided by the entropy density, or instead divided by the photon density. In this paper we adopt the entropy density normalization $L_6$, which is defined as:

\begin{equation}
    L_6 = 10^{6}\left(\dfrac{n-\bar n}{s}\right),
\end{equation}

\noindent where $n$ is the lepton density, $\bar n$ is the antilepton density, and $s$ is the entropy density. Results normalized in other studies by the photon density -- sometimes referred to as $L_4$ \citep{Abazajian14} and utilized in \citet{Bozek16} -- can be converted into $L_6$ equivalents through multiplying by $5.4\times10^{-3}$; we refer the reader to the appendix of \citet{Laine08} for a comprehensive discussion. 

The relationship of $\sim$~keV mass $N_1$ to the standard model provides a very different detection paradigm to that of other dark matter candidates. Constraints on the parameters of potential supersymmetric weakly interacting massive particles are set by non-detections at large annihilation interaction rates, and also depend on the complicated interactions that lead from the original annihilation to detectable products; lower bounds on the interaction are set from theoretical considerations \citep[see][for a review]{Gaskins16}. $N_1$ differ in that, for the range of \stt of interest for the 3.55~keV line, the `interaction rate' -- in this case the mixing angle between sterile neutrinos and standard model neutrinos -- correlates inversely with the mean velocity dispersion. The increase in velocity dispersion erases structures on progressively larger mass scales through free-streaming, and thus inhibits the formation of dwarf galaxies to a degree that is accessible to structure formation measurements. Therefore, the $N_1$ mixing angle is bounded by X-ray decay constraints from above and structure formation constraints from below in the 3.55~keV line region.   

The dominant decay channel of a sterile neutrino is to three active neutrinos \citep{Barger95}, and the subdominant X-ray decay channel is a two-body decay into an X-ray photon and a neutrino. In the latter case, conservation of momentum and energy determine that the photon energy will always be half the rest mass of the parent sterile neutrino, $m_\rmn{s}$, thus the mass will always be twice the rest-frame energy centroid of any measured line. The relationship between \stt and measured X-ray flux is relatively straightforward, and follows the relation:

\begin{equation}
    \Gamma_{\gamma}= 1.38\times10^{-29}\rmn{s^{-1}}\left(\frac{\sin^{2}(2\theta)}{10^{-7}}\right)\left(\frac{m_\rmn{s}}{1\rmn{keV}}\right)^{5},
\end{equation}

\noindent as discussed in \citet{Shrock74,Pal82,Barger95,Boyarsky14a} and \citet{Bulbul14}; $\Gamma_{\gamma}\approx\Gamma_{\nu\nu\nu}/128$ where $\Gamma_{\nu\nu\nu}$ is the decay rate in the dominant three-neutrino channel. \stt is also related to the production rate of dark matter in the early Universe. The maximum possible value of \stt at a given mass is set by the measured dark matter abundance, and this defines non-resonant production. The presence of a lepton asymmetry enhances the production rate as discussed above, which then requires a lower \stt. A lower limit on \stt is set by the maximum available lepton asymmetry, as determined by the number of available degrees of freedom. This limit has been identified as somewhere in the range $L_6=120$ to $L_6=700$  \citep{Boyarsky09a,Canetti13b,Canetti13a} and corresponds to \stt$\sim10^{-12}$; still lower values of \stt cannot generate the measured value of the dark matter abundance. For notionally higher values of $L_6$ a still more conservative result has been obtained from BBN constraints, which require $L_6\lsim2500$ \citep{Boyarsky09a,Cherry17} and therefore set \stt$>10^{-13}$.

In addition to the overall abundance of sterile neutrinos, the combination of mass, mixing angle, and lepton asymmetry also sets the sterile neutrino momentum distribution, and by extension the free-streaming scale, as mentioned above. A comprehensive discussion is presented in \citet{Lovell16}; here we provide a brief summary. Increasing $L_6$ promotes the resonant production of sterile neutrinos below some momentum threshold, and this momentum threshold itself increases with $L_6$. Therefore, decreasing \stt from its maximum value -- the non-resonant production value -- leads to a colder momentum distribution up to some characteristic value, beyond which the resonance momentum threshold is sufficiently high that the momentum distribution becomes warmer once again until eventually it is identical to the initial, non-resonant distribution. For the $m_\rmn{s}=7.1$~keV sterile neutrino the distribution is maximally cold, i.e. has the lowest mean velocity dispersion, at \stt$\sim2\times10^{-10}$. This is also the upper limit on \stt expected from the reported M31 3.55~keV line detection when the maximum uncertainty on the M31 mass is applied \citep{Boyarsky14a}; therefore, we are able to treat the velocity dispersion and \stt as inversely correlated for 3.55~keV line-compliant $N_1$ in the $\nu$MSM as discussed above. 

The momentum distribution can then be supplied to a Boltzmann solver code, such as {\sc camb} \citep{Lewis2000} or {\sc class} \citep{Lesgourgues11} to obtain a matter power spectrum, which in turn can be fed into cosmological simulations and semi-analytic models to estimate structure formation constraints. These constraints are typically described with parameters that apply to dark matter models beyond sterile neutrinos, such as the length-scale, specifically the half-mode wavenumber, $k_\rmn{hm}$. This is defined as the wavenumber at which the square root of the ratio of the considered dark matter power spectrum, labelled $P(k)$, relative to the power spectrum of cold dark matter (CDM), labelled $P_\rmn{CDM}(k)$, is 50~per~cent, or $\sqrt{P(k_\rmn{hm})/P_\rmn{CDM}(k_\rmn{hm})}=0.5$. This length-scale can be expressed as a mass-scale is known as the half-mode mass, $M_\rmn{hm}$ \citep{Viel13,Bose16a}, and is defined as

\begin{equation}
M_\rmn{hm} = \frac{4\upi}{3}\bar\rho\left(\frac{\upi}{k_\rmn{hm}}\right)^{3},
\end{equation}

\noindent where $\bar\rho$ is the average matter density of the Universe. Another choice is the thermal relic mass $m_\rmn{th}$, which is the mass of a notional thermal relic particle with a matter power spectrum that in some way approximates the dark matter model in question. In summary, X-ray detections and non-detections inform the permitted values of $m_\rmn{s}$ and \stt, and subsequent work is required to estimate the value of $L_6$ and also $M_\rmn{hm}$ for the purpose of obtaining astrophysics/cosmology constraints.

In practice, the ability to compute the lepton asymmetry and momentum distribution from the dark matter abundance, $m_\rmn{s}$ and \stt is non-trivial. Tracking the generation of sterile neutrinos from energies as high as $100$~GeV through to 10~MeV -- where production ceases -- involves computing the distribution of energies across the boundary where the quark--gluon plasma condenses into bound baryons, at around 150~MeV \citep{Abazajian01a,Petreczky12}, and also requires the computation of short-lived resonances in the lepton sector. \citet{Ghiglieri15} (hereafter \citetalias{Ghiglieri15}) and \citet{Venumadhav16} (hereafter \citetalias{Venumadhav16}) released public codes with the goal of computing these quantities, and demonstrated that differences in the modelling of neutrino interactions can have a large impact on the required $L_6$ at fixed \stt and $m_\rmn{s}$ -- \citetalias{Ghiglieri15} alone showed differences of a factor of 10 in the six cases it considered -- and that they will also have an impact on the matter power spectrum.

\subsection{Summary of time-scales and additional sterile neutrino models}

\begin{figure*}
    \includegraphics[scale=0.70]{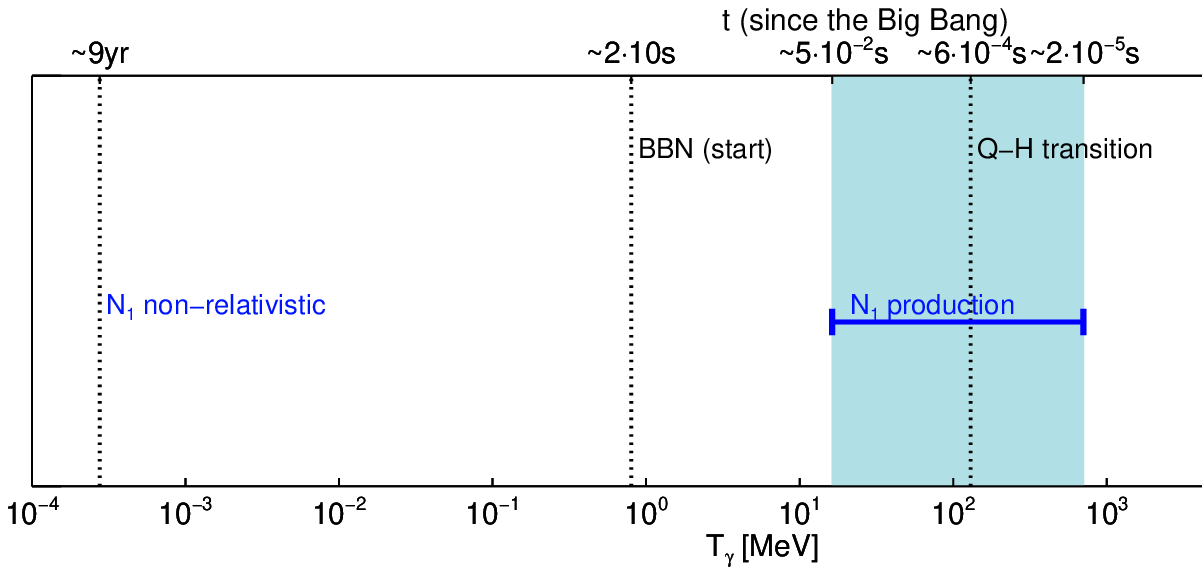}
    \caption{Rough illustration of energy scales associated with the production of $N_{1}$, where the energy scales are denoted with the photon temperature $T_{\gamma}$. The labelled energy scales are the start of BBN, the quark--hadron transition, the end of the sphaleron process, the epoch of $N_{1}$ production and the energy at which $N_{1}$ become non-relativistic. Approximate time-scales are indicated on the upper $x$-axis.}
    \label{fig:n1t}
\end{figure*}

We summarize some of the key points above by computing rough estimates of the temperatures and time-scales, and present these in Fig.~\ref{fig:n1t}. We indicate the scale at which sphaleron processes come to an end -- $T\sim10^{5}$~MeV and $t\sim10^{-10}$~s -- which also marks the closure of low-scale baryogenesis in the $\nu$MSM; the approximate quark--hadron transition at just above $10^{2}$~MeV and the start of BBN at 0.8~MeV. We then show the production time of $N_1$ as computed from \citetalias{Venumadhav16}, which occurs between $10^{-5}$ and $10^{-2}$~s after the Big Bang, and also the approximate time at which $N_1$ become non-relativistic, we calculate to be $\sim$9~yr after the Big Bang for \stt$=2\times10^{-11}$ and a slightly shorter period for cooler models.

This time of 9~yr is well before matter-radiation equality, which is $\gsim$50~000~yr after the Big Bang. Therefore, it is the case that the $N_1$ are relativistic during BBN. In principle, dark matter that is relativistic during BBN leads to a faster expansion rate and the injection of energy through annihilations, with the consequence that the helium fraction is drastically altered. Such considerations have ruled out thermal relic particles at the keV scale \citep{Depta19,Sabti20}. By contrast $N_1$ are completely inert and have no direct, significant interactions with the baryons at this epoch; they therefore evade these bounds. We have also performed a first order calculation of the change to the time period over which neutron decay is possible and thus the helium fraction is set: we find that the density parameter of $N_1$ at this time is $\Omega_\rmn{N1}\sim7\times10^{-7}$, and the time period of neutron decay is the same as non-relativistic dark matter to less than one part in a million \citep[see also][]{Dodelson94,Abazajian01a}. We assume for the purposes of this study that $N_1$ does not interfere with BBN, and defer a more careful study to future work. 

The most obvious challenge for this model to overcome is the opportunity for X-ray decay and structure formation constraints, which we discuss in Section~\ref{sec:obs}. An alternative method of ruling out this model is to instead make a positive detection of sterile neutrinos that have different parameters to those of the $\nu$MSM. The most prominent example is the light sterile neutrino, which is proposed to have a mass at the eV scale and to explain anomalies measured in the LSND and MiniBOONE experiments \citep{Athanassopoulos96,Mention11} but would not be a dark matter candidate; see the review of \citet{Abazajian17} for a further discussion of this topic. Recent analyses by the OPERA and MicroBOONE experiments have determined that these anomalies are in fact not present in their data \citep{Agafonova23,Abratenko23}. Therefore, the likelihood of the existence of the light sterile neutrino is diminished and thus this challenge to the $\nu$MSM version of sterile neutrino physics is no longer as salient. Further alternatives include sterile neutrinos generated from heavy scalars \citep{Kusenko06,Merle16}, TeV scale particles \citep{Humbert15} and production from light mediators of active neutrino interactions \citep{An23}, see \citet{Abazajian17} and \citet{Adhikari17} for a more complete list. In the case of decay detection, some of these options may produce the same result as the $\nu$MSM, in which case structure formation constraints will take on extra importance. The structure formation properties for these models are substantially different from the $\nu$MSM, and are described in \citet{Abazajian19} and \citet{Zelko22}.

In conclusion, we have discussed the background and merits of $N_1$ as a dark matter candidate, and outlined the challenges faced in estimating model parameters. In Section~\ref{sec:res} we will apply the \citetalias{Ghiglieri15} and \citetalias{Venumadhav16} codes to compute distribution functions and matter power spectra, illustrate the differences in these parameters between computational approaches, and compare the results to current constraints. 

\section{Computation of model parameters}
\label{sec:res}

We begin our analysis with the computation of a series of $N_1$ momentum distributions and their corresponding matter power spectra. We illustrate the differences between the predictions of different models, compare their results to the thermal relic model expectations and compute some key parameter values.

\subsection{Linear matter power spectra}

\begin{figure}
    \centering
    \includegraphics[scale=0.42]{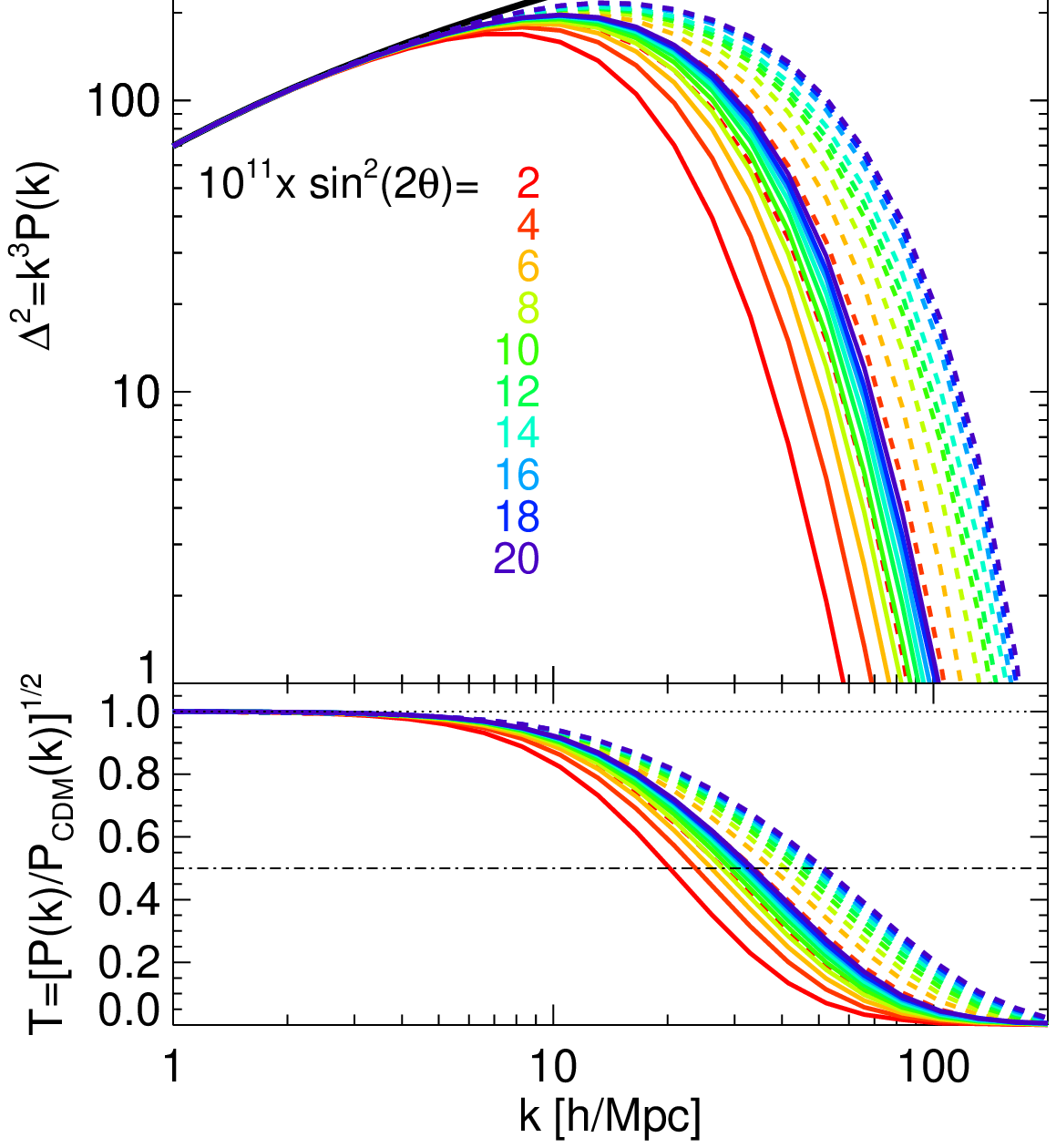}
    \caption{The dimensionless matter power spectra for dark matter models. $N_1$ power spectra mixing angles are indicated by their colour in the figure legend. \citetalias{Venumadhav16} spectra are indicated with solid lines and \citetalias{Ghiglieri15} with dashed lines. In the top panel CDM is indicated with a solid black panel; in the bottom panel we plot the transfer function of the $N_1$ power spectra with respect to CDM.}
    \label{fig:powspec}
\end{figure}

Our momentum distributions are computed as follows. We assume $m_\rmn{s}=7.1$~keV, as inspired by the 3.55~keV line, and adopt 10 values of \stt in the range $[2,20]\times10^{-11}$ to span the broadest range of viable \stt inferred from the 3.55~keV line detection of \citet{Boyarsky14a} in M31. We then compute momentum distributions for these ten mixing angles using the publicly available codes of \citetalias{Venumadhav16} and \citetalias{Ghiglieri15}. The latter code includes six different cases, labelled a--f: for all of the computations in this paper we adopt their case (d) mode as this the default non-equilibrium calculation presented in their online distribution. Also, this code determines cosmological abundance as a function of $L_6$ rather than the other way round, therefore we iterate over $L_6$ values to obtain an abundance that is within 1~per~cent of the measured cosmological dark matter abundance. We then compute matter power spectra using the {\sc class} \citep{Lesgourgues11} Boltzmann code, adopting the cosmological parameters determined in \citet{Planck16}. We plot the results for \citetalias{Venumadhav16} and \citetalias{Ghiglieri15} in Fig.~\ref{fig:powspec}. We have also performed these calculations with a modified version of {\sc camb} \citep{Lewis2000,Boyarsky09b} and obtain the same results as {\sc class} at the per~cent level in keeping with the findings of \citetalias{Venumadhav16}; we therefore use {\sc class} alone for the rest of this paper\footnote{ The original version of this paper used {\sc class} computations that significantly underestimated the power at small scales relative to {\sc camb}. We have corrected this error by running {\sc class} with the non-CDM fluid approximation switched off, as had been done previously in \citetalias{Venumadhav16}; this is achieved by setting {\tt ncdm\_fluid\_approximation=3} in the precision file.}.

Both codes return the familiar pattern in which the matter power spectrum exhibits a cutoff, and the wavenumber of the cutoff increases with \stt\footnote{Still larger values of \stt would exhibit a turnaround in behaviour, where the cutoff shifts back to smaller wavenumbers as discussed in the previous section. However, these are outside the expected 3.55~keV line constraints and therefore are not considered here.}. Where the two codes differ is in the location of the cutoff at fixed \stt: the \citetalias{Ghiglieri15}  turnovers are at consistently larger $k$ than for \citetalias{Venumadhav16}. This behaviour is more clearly apparent in the ratio with respect to CDM as shown in the bottom panel of Figure~\ref{fig:powspec}. It is apparent from this panel that the half mode wavenumber is marginally larger for \citetalias{Ghiglieri15} than it is for \citetalias{Venumadhav16}, thus the value of $M_\rmn{hm}$ will be smaller for the former. 

\begin{figure}
    \centering
    \includegraphics[scale=0.33]{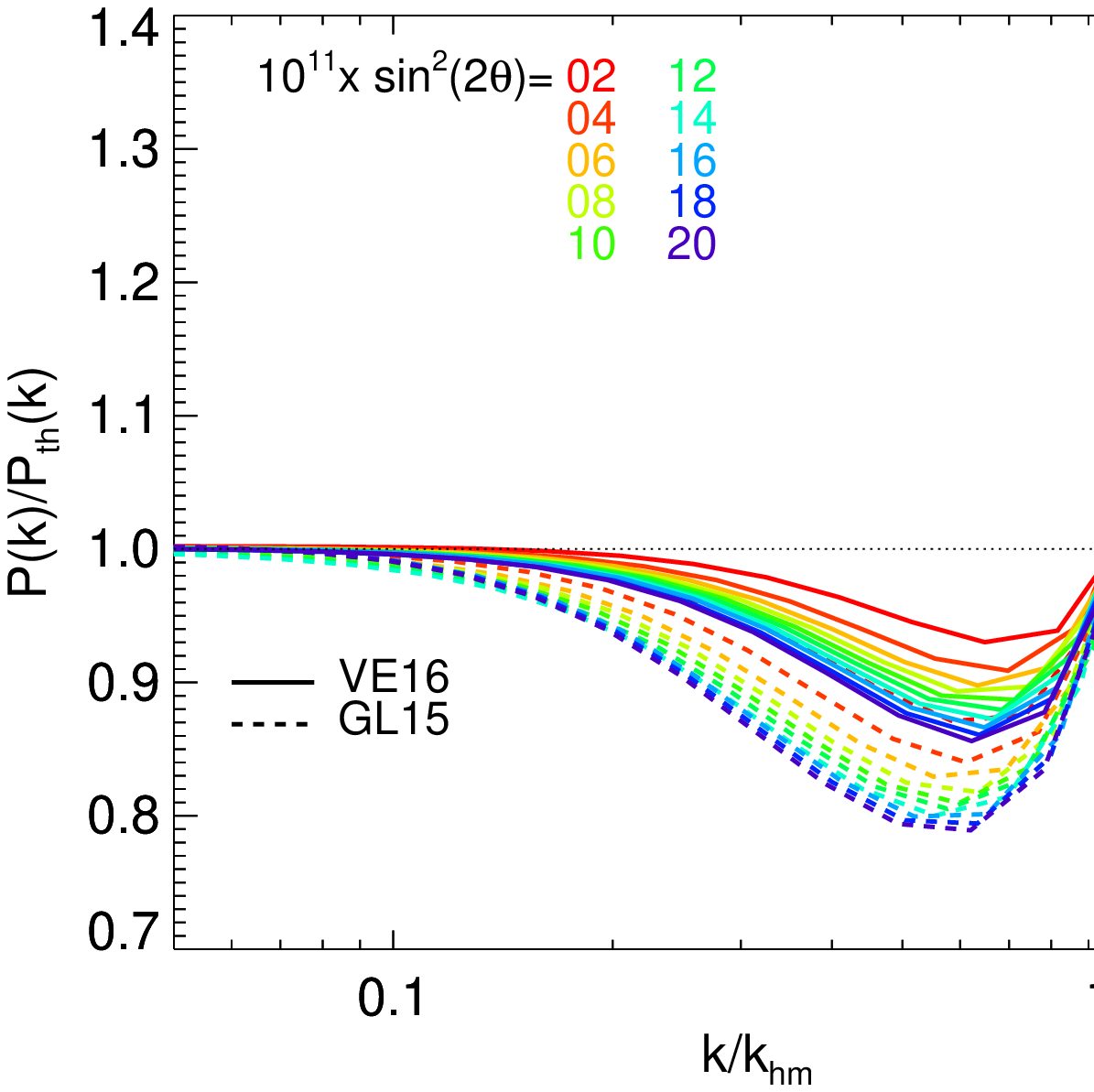}
    \caption{The ratio of the $N_1$ matter power spectra with respect to the \citet{Viel05} thermal relic approximation that has the same half-mode mass / half-mode wavenumber. Mixing angles are indicated by their colour in the figure legend.\citetalias{Venumadhav16} spectra are indicated with solid lines and \citetalias{Ghiglieri15} as dashed lines.}
    \label{fig:powspec_otr}
\end{figure}

The existence of a cutoff at scales corresponding to dwarf galaxies means that $N_1$ are classed as warm dark matter \citep[WDM; ][]{Bond83,Bardeen1986,Bode01}. Given the desire to have constraints on dark matter cutoffs that are not dependent on any single particle physics candidate, it is common to generate constraints for the thermal relic toy model discussed above. In practice, this is typically achieved through the fit to thermal relic spectra computed in \citet{Viel05}, which links the thermal relic mass $m_\rmn{th}$ to $M_\rmn{hm}$ as outlined in \citet{Bose16a}.\footnote{Subsequent work by \citet{Vogel23} has improved upon this fit; however, in cosmology and astrophysics studies it is the \citet{Viel05} fit that has been used to compute generic WDM limits and therefore we retain their equations in our work.}. The \citet{Viel05} relation in principle does not need to have the same shape as the $N_1$ spectra. Therefore, for each of our $N_1$ curves we compute the half-mode mass $M_\rmn{hm}$ and then in turn compute the \citet{Viel05} power spectrum that has the same $M_\rmn{hm}$. We compute the ratio of each $N_1$ power spectrum to its same-$M_\rmn{hm}$ thermal relic counterpart and present the results in Fig.~\ref{fig:powspec_otr}. We divide the wavenumber by each models' half-mode wavenumber, therefore the value of the ratio at $k/k_\rmn{hm}$ equals 1 by construction. 

The shallow cutoffs associated with both models significantly reduce the power with respect to the thermal relic approximation for $k<k_\rmn{hm}$ and exhibit more power than the thermal relic at larger wavenumbers. The difference between $N_1$ and the approximation is strongly dependent on \stt, with stronger deviations for larger values of \stt. The \citetalias{Ghiglieri15} \stt$=2\times10^{-10}$ model power is suppressed by 20~per~cent compared to the thermal relic equivalent compared to 12~per~cent for the same code with \stt$=2\times10^{-11}$. The difference is smaller for \citetalias{Venumadhav16}, at 10~per~cent for \stt$=2\times10^{-10}$ and smaller suppressions for lower \stt values. We therefore expect that the \citet{Viel05} thermal relic approximation overestimates the power associated with $N_1$ models, especially for the \citetalias{Ghiglieri15} momentum computations.

\subsection{Parameter values: lepton asymmetry and the half-mode mass}

\begin{figure*}
    \centering
    \includegraphics[scale=0.35]{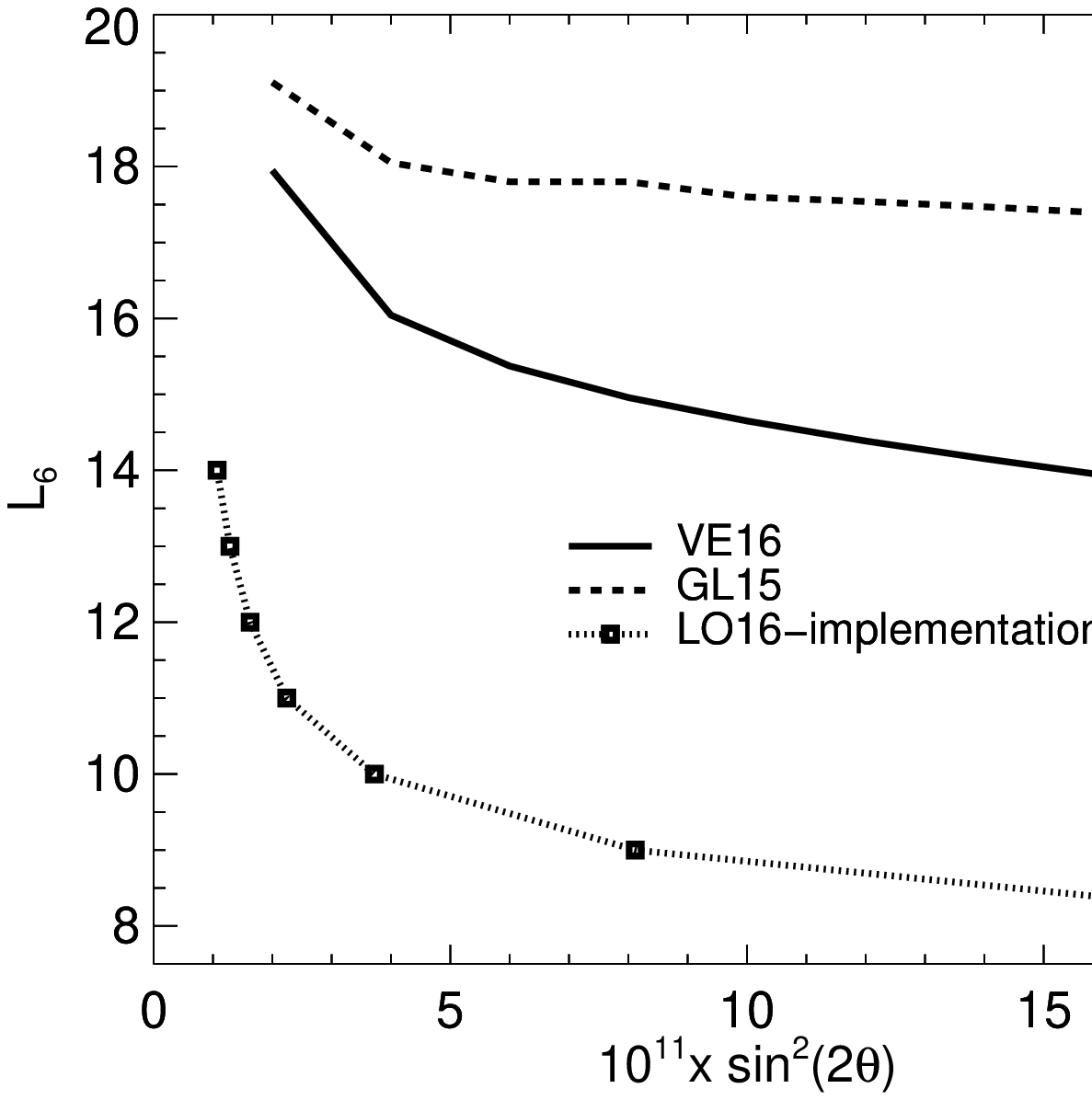}
    \includegraphics[scale=0.35]{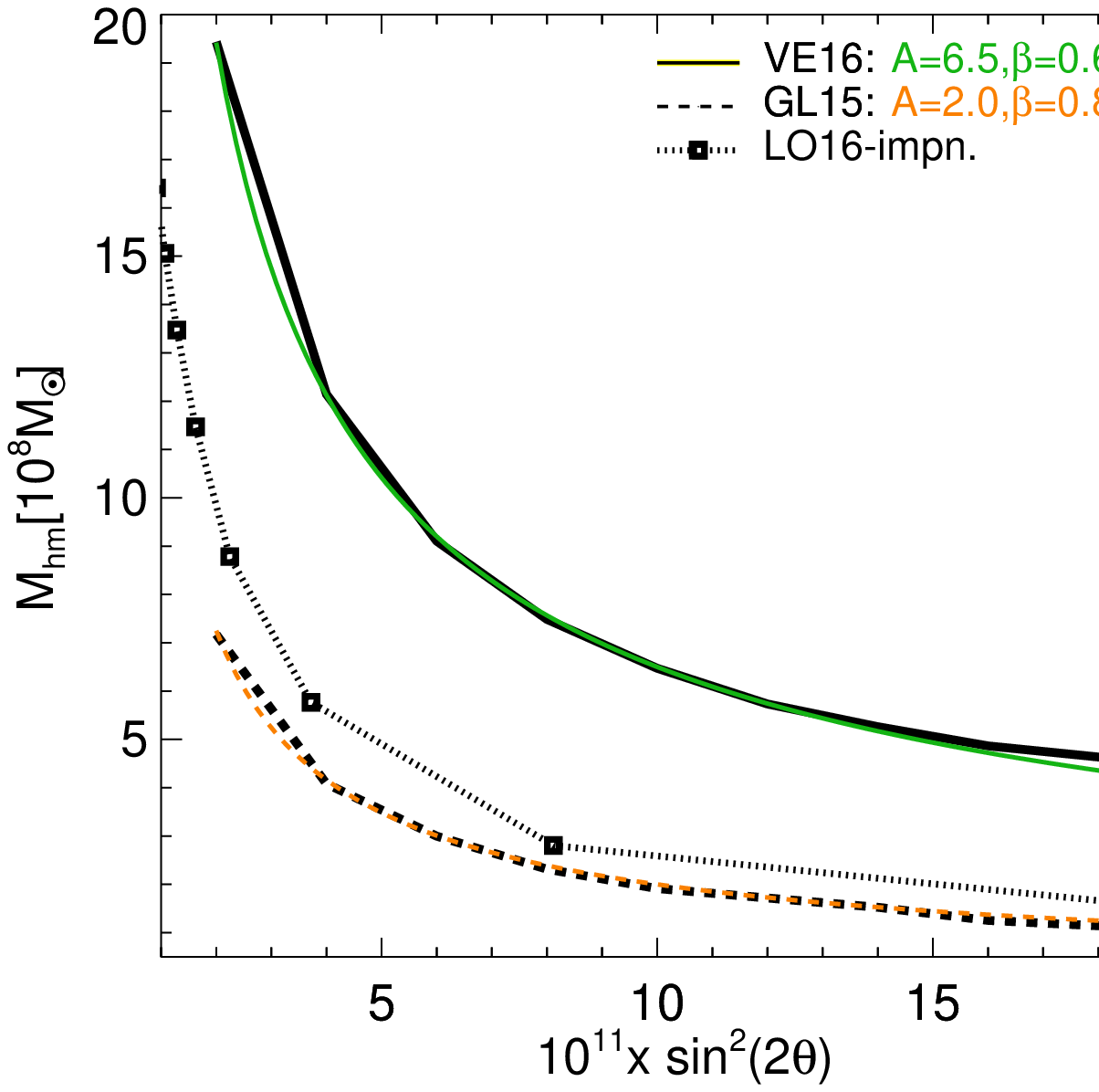}
    \caption{Model properties as a function of the mixing angle for different computational implementations. Left-hand panel: lepton asymmetry $L_6$. Right-hand panel: half-mode mass $M_\rmn{hm}$ with the corresponding thermal relic particle mass $m_\rmn{th}$ indicated on the right=hand $y$-axis. In both panels \citetalias{Venumadhav16} results are indicated with solid lines, \citetalias{Ghiglieri15} results are shown with dashed lines, and computations based on the \citet{Lovell16} implementation with dotted lines; in the latter case we show the individual data points with empty squares. In the right-hand panel we show fits to the  \citetalias{Venumadhav16} \citepalias{Ghiglieri15} results with green (orange) lines. We indicate the fitting parameters for equation~\ref{eqn:mhmfit} in the figure legend.}
    \label{fig:ps2}
\end{figure*}

We have demonstrated that the momentum distribution computation algorithm and parameter choices have a significant impact on the position of the cutoff for fixed \stt and $m_\rmn{s}$, and will therefore have an impact on $M_\rmn{hm}$. We also stated in Section~\ref{sec:rpsndm} that the value of $L_6$ required to obtain the correct abundance will also differ substantially. We illustrate the scale of these differences explicitly by computing $L_6$ and $M_\rmn{hm}$ as a function of \stt for both codes, and then presenting the results in Fig.~\ref{fig:ps2}. We also include the results of previously unpublished computations for the $7.1$~keV $N_1$ using the machinery behind \citet{Lovell16} and subsequent papers that build on that paper, which itself used an earlier version of the \citetalias{Ghiglieri15} code. We include these results to facilitate comparisons to earlier work; understanding the origins of these differences is beyond the scope of this paper.

The values of $L_6$ differ significantly between the three computational implementations. All three follow the familiar track wherein smaller mixing angles require larger lepton asymmetries to obtain the correct abundance, but the amplitudes of the relations differ considerably. \citetalias{Venumadhav16} prefers $L_6=18$ at \stt$=2\times10^{-11}$ down to $L_6=13.5$ at \stt$=2\times10^{-10}$, whereas the chosen implementation of \citetalias{Ghiglieri15} prefers $L_6=19$ to $L_6=17$ over the same range. We stress that this result applies to one specific implementation of \citetalias{Ghiglieri15}: alternative choices for the initial asymmetries of separate lepton species can generate overall asymmetries at \stt$=2\times10^{-11}$ that are as high as $L_6=102.77$ or as low as $L_6=13.47$, as discussed in \citetalias{Ghiglieri15} table~1. This lower value is somewhat closer to the \citet{Lovell16} implementation, which otherwise requires $L_6$ values much lower than either \citetalias{Venumadhav16} or our adopted \citetalias{Ghiglieri15} implementation. Therefore, there remains work to be done to estimate the initial lepton asymmetry precisely and accurately.

While the value of $L_6$ is inaccessible to current experiments, $M_\rmn{hm}$ is regularly subject to constraints. Both momentum codes return the familiar anticorrelation between \stt and $M_\rmn{hm}$, such that attempts to evade X-ray constraints on \stt lead to stronger structure formation constraints on $M_\rmn{hm}$ and vice versa. The amplitude of the relations is significantly different, with the smaller scale power spectrum cutoffs of \citetalias{Ghiglieri15} discussed in \ref{fig:powspec} leading to smaller $M_\rmn{hm}$ by a factor of $2$-$3$ than for \citetalias{Venumadhav16} at fixed \stt. The \citet{Lovell16} implementation generates a somewhat steeper relation, with $M_\rmn{hm}$ some 50~per~cent larger than \citetalias{Ghiglieri15} at \stt$=2\times10^{-11}$ but approaches the \citetalias{Ghiglieri15} values towards higher \stt. Finally, we generate fits to both implementations of the form

\begin{equation}
    M_\rmn{hm} = A\cdot10^{8}\msun\left(10^{11}\times\sin^{2}(2\theta)\right)^{-\beta},
    \label{eqn:mhmfit}
\end{equation}

\noindent where for \citetalias{Venumadhav16} $A=6.5$, $\beta=0.68$; and for \citetalias{Ghiglieri15} $A=2.0$, $\beta=0.80$.

\section{Comparison to observations}
\label{sec:obs}

In this section we assess the current status of X-ray detections and constraints on \stt, and of structure formation constraints on $M_\rmn{hm}$. We then compare some of these constraints to our fits for \citetalias{Venumadhav16} and \citetalias{Ghiglieri15}.

\subsection{X-ray detections and constraints}

The first reported detections of an X-ray line at 3.55~keV consistent with dark matter decay were made by \citet{Bulbul14} in clusters of galaxies with the {\it XMM-Newton} (`{\it XMM}') observatory and \citet{Boyarsky14a} in the M31 galaxy and the Perseus galaxy cluster, also with {\it XMM}. The values of \stt were estimated as $6.8^{+1.4}_{-1.4}\times10^{-11}$ ($1\sigma$ errors) for the cluster stack and $4.9^{+1.6}_{-1.3}\times10^{-11}$ for M31 on flux statistical uncertainties alone; including M31 mass uncertainties gives $[2-20]\times10^{-11}$. The detection in the centre of Perseus was anomalously high compared to the other results and subsequently ruled out by \citet{HitomiC17}. Further detections have been reported in the MW Galactic bulge at $\sim1^{\circ}$ from the Galactic Centre with {\it Chandra} \citep[$2.3\pm1.8\times10^{-11}$,][]{Hofmann19} and the MW halo contribution to the COSMOS Legacy and Extended {\it Chandra} Deep Field South survey fields, also with {\it Chandra} \citep[$\sim0.5-4\times10^{-10}$,][]{Cappelluti17}. Other studies have reported constraints that are in strong tension with the claimed detections, including in alternative studies of the Galactic Centre \citep{Jeltema14}, the broader MW halo (\citealp{Dessert20}, see also \citealp{Boyarsky20, Abazajian20a,Dessert20b}) and still further studies allow for only a small region of the parameter space of interest to remain \citep{Sicilian22,Roach23}. A stack of galaxies returned strong constraints \citep{Anderson14}, which motivated exotic dark matter models in which the line was only generated in clusters \citep{Conlon15}. Arguably the most promising target for a detection is the Draco dwarf spheroidal galaxy \citep{Lovell15}, yet 1.4~Ms observation with {\emph XMM} was insufficient to obtain a detection, and two different groups returned different results \citep{Jeltema16,Ruchayskiy16}.

Much of the above uncertainty reflects different approaches to modelling the astrophysical background. An alternative source for the line is charge exchange, in which electrons are accreted from the neutral medium onto sulphur atoms and then generate 3.55~keV photons through the subsequent cascade of atomic transitions \citep{Gu15,Shah16}, resulting in an astrophysical line. One key discriminator between a dark matter decay line and an astrophysical line is its velocity dispersion. Astrophysical line velocity dispersions are typically $<200$~$\kms$, while \citet{Lovell19c} showed that for nearby clusters the velocity dispersion is instead $450$-$800$~$\kms$ depending on the cluster. This measurement requires a high resolution X-ray calorimeter not available to the galaxy cluster studies discussed above. It had been hoped that the {\it Hitomi} mission would make the required measurement with its state-of-the-art calorimeter, but the mission was lost after a month and the data obtained prior to the failure were not sufficient to detect the line feature \citep{HitomiC17}. The next opportunity to test the line velocity dispersion will come with the {\it XRISM} mission \citep{XRISM21}.

\subsection{Structure formation and cosmological constraints}

Cosmological and structure formation constraints have been derived in a wide variety of astronomical observables, including the CMB, BBN, the Lyman-$\alpha$ forest, gravitational lensing, stellar streams around the MW, reionization studies and MW satellite counts. We summarize the status of each of these fields below. Many of the cosmological constraints are defined for a generic thermal relic WDM model. They typically assume the WDM thermal relic power spectrum using the $m_\rmn{th}$ parameter; recall that this reflects an approximation to the $N_1$ power spectrum and is not the same as the $N_1$ mass. Throughout this section we will refer to constraints on the WDM model rather than $N_1$, except where noted otherwise. 

Part of the challenge in setting limits through these studies is the difficulty in selecting a lower effective prior on the parameters. $m_\rmn{th}$ of viable true thermal relic models is $\gsim100$~MeV \citep{Depta19,Sabti20} and is CDM, whereas the effective $m_\rmn{th}$ for a classical axion model may well be much larger than this, and all at much larger values than the 1-10~keV expected for WDM. One approach is to instead constraint $1/m_\rmn{th}$, which for CDM $=0$~$\rmn{keV}^{-1}$. This option evades the infinity but still creates a problem in logarithmic limits, a problem that is shared by $M_\rmn{hm}$. Most studies typically quote a limit on each parameter at 95~per~cent confidence level (95~C.~L.), based on the probability distribution of a posterior likelihood. Some studies also quote the highest $M_\rmn{hm}$ (lowest $m_\rmn{th}$) for which the likelihood is 5~per~cent of the maximum likelihood amplitude (5~M.~L.). The latter measure is more conservative and evades the issue of how to determine the lower end of the prior. Therefore, we will set a preference to use this option in what follows, including inferring this result from paper results where possible.  

\subsubsection{Early Universe constraints} 

If $N_2$ and $N_3$ are sufficiently long-lived, with lifetime $\tau>0.1$~s, they will interfere significantly with BBN, it is thus possible to show that $L_6<2500$ \citep{Serpico05}. Moreover, it is conceivable that if a sufficient fraction of the sterile neutrinos remain relativistic to late times, they may add a contribution to the number of relativistic degrees of freedom, $N_\rmn{eff}$. Our {\sc class} calculations return $\Delta N_\rmn{eff}<1.1\times10^{-3}$ for all \citetalias{Venumadhav16} and \citetalias{Ghiglieri15} models, which is significantly smaller than the $\sim0.35$ figure (95~C.~L.) of \citet{PlanckCP20}. We present these results in more detail in Appendix~\ref{app:neff}.  

\subsubsection{The Lyman-$\alpha$ forest}
Lyman-$\alpha$ constraints take advantage of the suppression of structure at redshifts $4<z<6$. They examine fine hydrogen absorption features in the spectra of high redshift quasars -- the Lyman-$\alpha$ forest -- to infer the amount of structure present in the gas, and convolve these results with expectations for the properties of gas physics to obtain the underlying dark matter distribution. Progressively stronger constraints have been reported, from $m_\rmn{th}>3.3$~keV \citep{Viel05} at 95~C.~L. through to $m_\rmn{th}>5.2$~keV (95~C.~L.) \citep{Irsic17}  (see also \citealp{murgia2017}); we infer $m_\rmn{th}>4.7$~keV 5~M.~L. for these data. Arguably the primary source of uncertainty is the thermal history, which may either mimic or obscure a dark matter power spectrum cutoff \citep{Garzilli17}. A recent analysis by \citet{Villasenor23} used new data sets and a different approach to the thermal history. They set a limit of $>3.1$~keV (95~C.~L.; we infer $>3.0$~keV for 5~M.~L.) and reported a weak preference for $m_\rmn{th}=4.5^{+45}_{-1.4}$~keV over the CDM model.

\subsubsection{\it Gravitation lensing}
While the Lyman~$\alpha$ forest studies use a background light source to detect a power spectrum cutoff through absorption, gravitational lensing experiments instead infer dark matter properties from how the background source is either distorted or magnified. The background light is lensed by a foreground dark matter halo -- typically an isolated elliptical galaxy -- and the degree of dark matter substructure is determined by the degree to which the lensing signal deviates from that expected for a smooth lens. This method has the potential to detect dark matter haloes down to $10^{6}$~$\msun$, where haloes are expected to contain no gas and thus remain dark. The two subbranches of lensing relevant for linear matter power spectrum studies are flux anomalies, in which the background source is a multiply imaged quasar, and gravitational imaging, which instead analyses lensed images of extended galaxies.

The gravitational lensing method has successfully detected several massive ($>10^{9}$~$\msun$) haloes \citep{Vegetti10,Hezaveh16}, but not to a degree sufficient to compute strong constraints on the WDM model \citep{Vegetti18,Ritondale18}, and Euclid and the Vera C. Rubin Observatory (VCRO) will not have enough angular resolution to improve the constraints \citep{ORiordan23}. It is expected that high resolution imaging from the {\it European Extremely Large Telescope} ({\it E-ELT}) or very large baseline interferometry coupled to the square kilometre-array \citep[][]{McKean15,Powell23} will be required to obtain sufficient statistics. Recent work in the flux anomalies side has claimed stronger constraints: \citet{Gilman20a} reports $m_\rmn{th}>5.2$~keV 95~C.~L. ($>3.2$~keV at 5~M.~L., see \citealp{Zelko22} for an application of this result to sterile neutrino models), and \citet{Hsueh19} find $m_\rmn{th}>5.58$~keV 95~C.~L.,\footnote{We do not infer a 5~M.~L. limit for this result because the joint posterior amplitude is higher than 5~per~cent of its maximum at all quoted masses.} although the latter also reports the role of systematic uncertainties that require further study to refine this constraint. The {\it Euclid} mission may detect enough flux anomalies systems to discern whether the flux anomaly distribution is better described by CDM or WDM \citep{Harvey20}.

\subsubsection{Stellar stream gaps}
An alternative method for detecting dark subhaloes is to search for their impact on gaps in stellar streams up to $40$~kpc from the MW centre. Stellar streams are generated through the tidal stripping of globular clusters and satellite galaxies by the MW, and are expected to be very smooth in their light profile. The detection of kinks or gaps in a stream is then evidence for a past interaction with a perturber, including dark matter subhaloes. One reported detection of such a perturbation was made by \citet[][see also \citealp{Ibata20}]{Bonaca19}. \citet{Banik21b} reported $m_\rmn{th}>4.9$~keV (95~C.~L.) based on expectations for the MW subhalo mass function in WDM. However, the subhalo mass function in the central $<40$~kpc differs less from CDM than is the case for the halo as a whole \citep{Lovell21}; taking this factor into account weakens the constraint to $m_\rmn{th}>2.1$~keV (95~C.~L.).

\subsubsection{\it High redshift constraints}
The previous three types of studies looked for gravitationally induced distortions on observables. One further approach is to detect light from dwarf galaxies, whose number density is suppressed in WDM models. One such source of light is low mass galaxies in the early Universe, the presence or absence of which impacts the rate of reionization and thus determines when the Universe becomes transparent to visible light/mildly opaque to CMB photons. Various studies have considered the possibility of detecting these galaxies directly or instead comparing to reionization constraints. This possibility has been considered using a variety of simulation- and semi-analytic model-based approaches \citep{Schultz14,Bose16c,Menci16,Rudakovskyi16,Lovell19b,Rudakovskyi21,Kurmus22,Maio23}. These papers reveal that the impact of the cutoff is highly degenerate with gas physics processes such as the escape fraction of photons, and are therefore results in constraints weaker than those of the other observables listed above. The reported limits include $m_\rmn{th}>2.4$~keV \citep[95~C.~L., ][]{Menci16}, and $m_\rmn{th}>2$~keV \citep{Rudakovskyi16,Maio23}. 

\subsubsection{\it MW satellite counts}
The final method that we will consider here for generating constraints is the detection of MW satellite galaxies. The MW satellites are the faintest observable galaxies, and therefore probe the edge of reionization as well as the subhalo mass function at $\sim10^{9}$~$\msun$. Constraints rely on estimating the true number of MW satellites: approximately 50 are known \citep[e.g.][]{Bechtol15,DrlicaWagner15,Torrealba16,Koposov18} and the number of undetected satellites are inferred from completeness limits due to depth and sky coverage. Very conservative limits can be obtained for values of $m_\rmn{th}<1.6$~keV where the number of satellites generated is even smaller than the number of known MW satellites \citep{Lovell14} and marginally less conservative limits return $m_\rmn{th}<2.3$~keV \citep{Polisensky11}. 

Stronger limits require an understanding of the sky coverage selection function discussed above as well as physical processes such as the evacuation of gas from low mass haloes during reionization, the disruption of satellites by the MW disc, uncertainties in the MW halo mass, stochastic variations in halo assembly and the role of the Large Magellanic Cloud (LMC). Different assumptions in these areas can lead to dramatic differences in the number of predicted satellites: \citet{Cherry17} estimated $\sim100$ satellites, the compilation by \citet{Kim18} reports 120-150 satellites, the analysis by \citet{Newton18} expects $\sim120$ satellites, and \citet{Nadler20} instead predicts of order 220. These results are key inputs for further studies of both WDM generally \citep{Enzi21,Nadler21,Newton21} and $N_1$ \citep{Dekker22,Zelko22}, and we discuss these at the end of this subsection. 

\subsubsection{\it MW satellite structure}
We end this review on the interplay between the WDM model and astrophysical observations with a discussion of the structure of MW satellites. The density of haloes with masses around the cutoff scale is suppressed in WDM relative to CDM \citep{Colin00,Lovell12}, and the degree of suppression is sufficient to explain a claimed discrepancy in the masses of subhaloes in CDM simulations versus observed satellites known as the `Too Big To Fail' problem \citep{BoylanKolchin11,BoylanKolchin12,Lovell12,horiuchi2016,Lovell17a,Lovell17b,Bozek19,Lovell23a}. In principle, the density profiles of these satellites are the single most informative source for discriminating between different dark matter models: from high concentration cusps in CDM \citep{NFW_96,NFW_97} to lower concentration cusps in WDM \citep{Lovell14}; from self-interacting dark matter (SIDM) cores \citep{Vogelsberger12} to the very steep gravothermally collapsed cusps in extreme SIDM models \citep{Zavala19a} and even the solitons of fuzzy dark matter \citep{Nori23}. However, in the faintest systems it is very challenging to obtain enough stars to compute a density profile, and in brighter galaxies supernova feedback may play a confounding role \citep{Navarro_97,Pontzen_Governato_11}. Therefore, we do not attempt to provide WDM parameter constraints using satellite densities but highlight the importance of this observable for future dark matter studies. 

\subsubsection{Combined constraints}
Several of the studies discussed above have reported combined constraints that convolve posterior distributions from a number of different observations. These include \citet[][lensing, satellite counts, and Lyman-$\alpha$ forest]{Enzi21}, \citet[][lensing and satellite counts]{Nadler21} and \citet[][lensing, satellite counts, and Lyman-$\alpha$ forest]{Zelko22}. In all cases the strongest limits are provided by the satellite count component: therefore, as stated above for the two studies that constrain the \citet{Viel05} thermal relic model, $M_\rmn{hm}<8.5\times10^{8}$~$\msun$ (5~M.~L.) for \citet{Enzi21} and $M_\rmn{hm}<2.5\times10^{7}$~$\msun$ (5~M.~L.) for \citet{Nadler21}. As we will demonstrate below, the difference between these results brackets the allowed range for 3.55~keV line-compliant $N_1$.

\subsection{Comparison to sterile neutrino parameter relations}

In the previous two subsections we have presented limits from observations on $N_1$/WDM parameters. X-ray non-detections enforce lower values of \stt, and WDM structure formation arguments decrease the allowed $M_\rmn{hm}$. Given that \stt and $M_\rmn{hm}$ are anticorrelated, it is therefore possible to compare these in tension to estimate the optimal $N_1$ parameters, if any, that are permitted by the observations or even preferred by reported detections.

\begin{figure*}
    \centering
    \includegraphics[scale = 0.71]{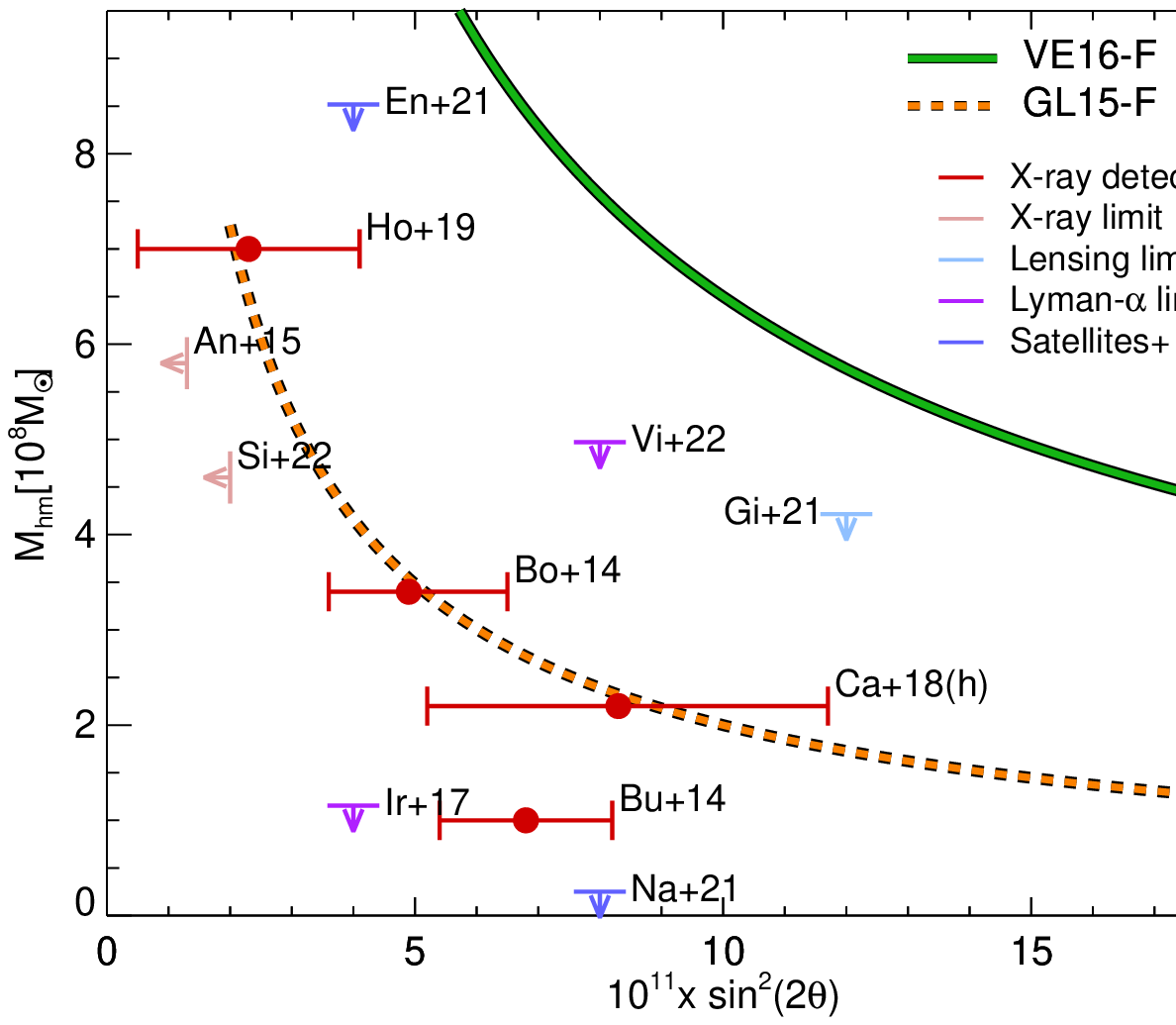}
\caption{Constraints on the half-mode mass $M_\rmn{hm}$ as a function of mixing angle with the corresponding thermal relic particle mass $m_\rmn{th}$ indicated on the right-hand $y$-axis. The fits to \citetalias{Venumadhav16} and \citetalias{Ghiglieri15} from Fig.~\ref{fig:ps2} are reproduced as solid green and dashed orange lines respectively. Reported 3.55~keV line detections are indicated as red error bars and reflect $1\sigma$ uncertainties. X-ray non-detection-derived upper limits on the mixing angle are indicated with pink vertical lines, and upper limits on $M_\rmn{hm}$ are shown with horizontal lines; the colours correspond to different types of observations as indicated in the figure legend. The $x$-axis location of $M_\rmn{hm}$ constraints is arbitrary, as is the $y$-axis location of the X-ray constraints and detections.}
\label{fig:cons}
\end{figure*}

In Fig.~\ref{fig:cons}, we compare some of the X-ray and structure formation constraints listed above to the equation~\ref{eqn:mhmfit} fits to \citetalias{Venumadhav16} and \citetalias{Ghiglieri15}. Upper limits on $M_\rmn{hm}$ are shown as horizontal lines with down-facing arrows, and upper limits on \stt from X-ray studies are shown vertical lines with left-facing arrows; X-ray detections are shown as $1\sigma$ error bars. Each constraint only applies to either $M_\rmn{hm}$ or \stt, and therefore its line applies across all of the presented parameter space; we draw short lines for the sake of clarity. We include four reported detections that span a range of targets: the M31 detection of \citet{Boyarsky14a}, the galaxy cluster stack of \citet{Bulbul14}, the Galactic bulge detection by \citet{Hofmann19}, and the outer MW halo detection of \citet{Cappelluti17} when adopting the high-mass MW halo parameters. We represent non-detections with two of the most stringent available: the galaxy stack of \citet{Anderson14} and the MW halo work of \citet{Sicilian22} \citep[for further analysis of the MW halo see][]{Dessert20,Boyarsky20,Dessert20b}. Finally, for $M_\rmn{hm}$ constraints we include five studies that quote, or from which we can obtain, 5~M.~L. limits: two Lyman-$\alpha$ results \citep{Irsic17,Villasenor23}, one gravitational lensing study \citep{Gilman20a} and two combined-observation analyses that are in practice dominated by MW satellite counts \citep{Enzi21,Nadler21}\footnote{Strong constraints have also been presented by \citet{Kim18} ($m_\rmn{th}>4$~keV, MW satellite counts) and \citet{Hsueh19} ($m_\rmn{th}>5.3$, 95~C.~L., gravitational lensing); it is not possible to obtain 5~M.~L. limits from these studies.}. 

In both the X-ray observations and the structure formation analyses, there is a wide variety of inconsistent constraints, some of which are in strong tension with both $N_1$ model predictions. The satellite counts in \citet{Nadler21} are incompatible with any part of the suggested $N_1$ parameter space, with $M_\rmn{hm}<5\times10^{7}$~$\msun$ whereas the alternative analysis by \citet{Enzi21} is consistent with all models but for the \citetalias{Venumadhav16} model with \stt$<7\times10^{-11}$. These studies are strongly dependent on the modelling of the LMC satellite contribution, the disruption of satellites by the disc, and the mass of MW halo, each of which leads to the differences in these results. There is similar disagreement within the Lyman-$\alpha$ bounds, where the \citet{Irsic17} limit rules out all $N_1$ models while \citet{Villasenor23} is consistent with \citetalias{Ghiglieri15} for \stt$>3\times10^{-11}$. The \citet{Gilman20a} limit is consistent with \citetalias{Ghiglieri15} for \stt$>4\times10^{-11}$ and rules out all of the \citetalias{Venumadhav16} options.

The X-ray results are similarly inconsistent. We quote two non-detection studies, \citet{Sicilian22} and \citet{Anderson14}, that are strongly constraining. The \citet{Sicilian22} result permits \stt$<2\times10^{-11}$, which is consistent with the \citet{Hofmann19} detection but in some tension with the other. It would also require $M_\rmn{hm}>7\times10^{8}$~$\msun$ for \citetalias{Ghiglieri15}. The \citet{Anderson14} analysis is instead inconsistent with all of the allowed $N_1$ range. As for the reported detections, there is no one value  of \stt that is consistent with the $1\sigma$ ranges of all four. Adopting the value of $5\times10^{-11}$ would represent a median between the four observations; if one instead posited that there is an undetected systematic uncertainty in the MW halo-derived results -- both for the \citet{Hofmann19} detection and for the limits by \citet{Dessert20,Dessert20b} and \citet{Sicilian22} -- that underestimates \stt then one can make a case for $6\times10^{-11}$. 

The question then becomes how to process these data into a coherent picture. The simplest approach is to take the most stringent constraints at face value, in which case both of the $N_1$ implementations discussed here are ruled out through a combination of the \citet{Anderson14} galaxy stack and the \citet{Nadler21} satellite counts. One then has to consider either that further iterations of the $N_1$ momentum distribution calculation will return values of $M_\rmn{hm}$ that are lower by at least a factor of 20 compared to \citetalias{Ghiglieri15}, in order to obtain $M_\rmn{hm}<5\times10^{7}$~$\msun$ at \stt$\sim1\times10^{-11}$, or that resonantly produced $m_\rmn{s}=7.1$~keV $N_1$s as a whole are not viable and should be discarded as a dark matter candidate. If one were to be more optimistic, one could attempt to determine a provisional value of \stt, first by comparing the detections and then factoring in those X-ray and structure formation constraints that do result in a coherent \stt--$M_\rmn{hm}$ pair. The rough mid-point of the reported detections would return \stt$\sim5\times10^{-11}$; relaxing the contribution of the MW halo could instead lead to $6\times10^{-11}$ and then $M_\rmn{hm}\sim3.0\times10^{8}$~$\msun$ using \citetalias{Ghiglieri15}. This choice still gives \stt $3\times$ the \citet{Sicilian22}  constraint and $M_\rmn{hm}$ $3\times$ higher than the \citet{Irsic17} constraint. Given that privileging one constraint will lead to greater tension with the other, we will argue that the single best value of \stt for future study is \stt$=6\times10^{-11}$, with $M_\rmn{hm}\sim3.0\times10^{8}$~$\msun$ -- and by extension $m_\rmn{th}\sim3.5$~keV -- as calculated with \citetalias{Ghiglieri15}. This value is also well within the preferred region for $m_\rmn{th}$ determined by \citet{Villasenor23}. We focus on this model for future study and for a comprehensive evaluation together with the strongest reported constraints.

\section{Conclusions}
\label{sec:conc}

The resonantly produced sterile neutrino ($N_1$) constitutes a compelling dark matter candidate for several reasons. It is part of a larger theory that may also explain neutrino oscillations and baryogenesis \citep{Dodelson94,Dolgov02,Asaka05,Laine08,Boyarsky09a}, as well as adding right-handed neutrinos to the standard model. It is also accessible to astronomical experiments, through its decay into X-rays for large mixing angles \citep{Pal82,Abazajian01a,Abazajian:01b} and its erasure of structure at small mixing angles \citep{Dodelson94,Shi99,Laine08,Lovell16}. Detections of an X-ray signal claimed to be compatible with the decay of a $N_1$ with mass $m_\rmn{s}=7.1$~keV and mixing angle \stt$=[2,20]\times10^{-11}$ add extra motivation. In this paper we revisit the physics behind $N_1$ production and its role in neutrino oscillations and baryons, present how different computational methods reach different conclusions for the linear matter power spectrum, and compare the results to observational constraints.

The $\nu$MSM proposes the addition of three sterile neutrinos to the standard model. If two of these sterile neutrinos have masses at the GeV scale ($N_2$ and $N_3$) and $N_1$ is at the keV scale, $N_2$ and $N_3$ can facilitate baryogenesis and neutrino oscillations while $N_1$ is a dark matter candidate. The same process that leads $N_2$ and $N_3$ to generate the baryon asymmetry of the Universe may also lead to a lepton asymmetry at later times, which then enables the resonant production of $N_1$. This resonant production leads to a momentum distribution that is skewed towards lower momenta than is the case for non-resonant production. The result is that the model behaves as WDM. 

Two different publicly available codes -- published in \citetalias{Venumadhav16} and \citetalias{Ghiglieri15} -- compute the momentum distributions for $N_1$, determining the correct value of the lepton asymmetry, $L_6$, to obtain the required dark matter abundance. We echo the result of \citetalias{Venumadhav16} in demonstrating that their code generates a warmer matter power spectrum than does \citetalias{Ghiglieri15} (Fig.~\ref{fig:powspec}). Sterile neutrino models are often approximated by thermal relic power spectra: we show that the thermal relic model overestimates the power in sterile neutrino models for $k<k_\rmn{hm}$ by up to 20~per~cent, especially for larger \stt (Fig.~\ref{fig:powspec_otr}). The value of $L_6$ for \citetalias{Venumadhav16} and the chosen implementation of \citetalias{Ghiglieri15} varies by 26~per~cent (Fig.~\ref{fig:ps2}), but further implementations will differ by a factor of 10 \citepalias{Ghiglieri15}. We also compute the half-mode mass $M_\rmn{hm}$ as a function of \stt, and show this variable can vary by up to a factor of 5 between the two implementations.

We discuss the current status of constraints of dark matter decay from X-rays and the limits on the WDM cutoff scale from structure formation measurements. We present four detections of the 3.55~keV line centred on \stt$\sim5\times10^{-11}$ and several non-detections that prefer \stt$<2\times10^{-11}$. We also discuss structure formation constraints on $M_\rmn{hm}$ from the Lyman-$\alpha$ forest, gravitational lensing, stream gaps, and MW satellite counts. We show that the satellite counts currently offer the strongest constraints, and competing analyses offer limits that bracket the $N_1$ $M_\rmn{hm}$ range: $M_\rmn{hm}<8.5\times10^{8}$~$\msun$ at 5~M.~L. for \citet{Enzi21} and $M_\rmn{hm}<3\times10^{7}$~$\msun$ at 5~M.~L. for \citet{Nadler21} (Fig.~\ref{fig:cons}). If we adopt the most stringent published constraints, then  $N_1$ dark matter is ruled out; if we instead take an optimistic approach then the combination of \stt and $M_\rmn{hm}$ constraints suggests the average detection \stt of $\sim6\times10^{-11}$ is most likely.

The clearest test of this result in the near future is the launch of the {\it XRISM} mission. This facility features a high resolution calorimeter that will measure the widths of X-ray lines, which in clusters of galaxies are characteristically larger for dark matter decay ($500-700$~$\kms$) than for astrophysical lines ($<200$~$\kms$). The performance verification (PV) phase for {\it XRISM} will include the following guaranteed observation time: 280~ks for the Perseus cluster -- comparable to the 230~ks obtained by Hitomi before that mission was lost -- plus 500~ks of the Virgo cluster, 150~ks of the Centaurus cluster and 200~ks of the Coma cluster, with an additional 200~ks for Perseus and 100~ks for Coma if conditions allow (\href{https://heasarc.gsfc.nasa.gov/docs/xrism/timelines/pvtargets.html}{https://heasarc.gsfc.nasa.gov/docs/xrism/timelines/pvtargets.html}). 

The Perseus, Virgo, and Centaurus decay line parameters were estimated for the {\it XRISM} field of view in \citet{Lovell19c}, with on-centre line-of-sight velocity dispersions $\sigma_\rmn{1D}=\sim600$, $\sim470$ and $\sim450$~$\kms$ respectively. The Coma cluster mass and distance parameters are approximately the same as Perseus and therefore this cluster should also have $\sigma_\rmn{1D}\sim600$~$\kms$. \citet{Lovell19c} also estimated that the expected flux from Virgo should be twice that of Perseus, thus when combined with the PV time allocations suggests that Virgo may present the first opportunity to detect the line, prior to the consideration of other observational effects. \citet{Bulbul14} estimated that a Perseus line of $\sigma_\rmn{1D}=1300$~$\kms$ -- the velocity dispersion of the Perseus member galaxies rather than the dark matter -- could be detected with 1~Ms of exposure, which is $>2\times$ the maximum Perseus PV allocation but to first order returns the same number of photons as the 500~ks Virgo allocation, given that the expected Virgo flux is twice that of Perseus and also that the line width in Virgo is half that of the \citet{Bulbul14}-Perseus estimate. The tentative detection of a line in Virgo would then motivate dedicated, post-PV campaigns to measure the line width in other nearby clusters and ascertain whether or not its properties strictly adhere to the dark matter content.    

In the event that such a detection is made, two questions will arise: (i) whether the line is indeed dark matter decay, and (ii) whether this candidate is $N_1$ dark matter. First, the following steps will need to be taken to ascertain that the line is dark matter decay: 

\begin{itemize}
    \item Obtain careful estimates of the astrophysical background signal in galaxies and galaxy clusters, empirically from observations and also from mock observations of simulations.
    \item Perform observations of targets where the amount of hot gas is minimized, such as the Bullet cluster \citep{Boyarsky08a} and the Draco dSph \citep{Lovell15}.
    \item Develop sophisticated models of the MW halo mass distribution, including adiabatic contraction, to compare to blank sky X-ray observations.
\end{itemize}

Second, we set out the issues to be addressed in ascertaining whether this decaying dark matter is $N_1$ rather than a still more exotic candidate:

\begin{itemize}
    \item Refine the computations of early Universe physics to ascertain as accurately as possible the initial lepton asymmetry and the momentum distribution.
    \item Identify whether differences between the full $N_1$ matter power spectra calculations and the thermal relic approximation are significant for structure formation constraints.
    \item Ascertain the source of the current incompatibility in MW satellite counts, and then resolve the discrepancy, including through new observations by the upcoming {\it VCRO-LSST} facility \citep{LSST19}.  
    \item Perform high-resolution gravitational imaging studies with the E-ELT to measure the $\sim10^{8}$~$\msun$ halo mass function \citep{Vegetti18}.
    \item Continue to develop models of galaxy formation that accurately follow processes as diverse as star formation, supernova feedback, magnetic fields, and stripping by host galaxies in order to predict halo density profiles and luminosity fractions, particularly in dSphs \citep{Gutcke22}.
    \item Pursue efforts to detect other candidates, such as low-reheating temperature models where the signal is explained by a particle that is only a small fraction of the dark matter \citep{Abazajian17} and could be experimentally detected by the proposed HUNTER experiment \citep{Martoff21}.
\end{itemize}

One final consideration is the contribution from particle physics experiments. Generating measurable amount of $N_1$ in an experiment is a formidable challenge, where even the anticipated upgrades to the KATRIN experiment are limited to probing \stt$>10^{-9}$ \citep{KATRIN22}, some two orders of magnitude higher than the lowest 3.55~keV line-compatible models. Other experiments are instead searching for $N_2$ and $N_3$. The DUNE experiment will probe neutrino physics to ascertain the neutrino mass hierarchy and measure the charge-parity-violating phase, and has some potential to constrain the parameter space of $N_2$ and $N_3$ \citep{DUNE15,DUNE20}; it will come online in the second half of this decade. Two proposed CERN experiments, SHiP \citep{SHiP19} and MATHUSLA \citep{MATHUSLA19}, would probe rare events in which $N_2$ and $N_3$ are produced and then decay into products that are subsequently detected. The combination of experiment, X-ray observations, and structure formation constraints will then have played a key role in identifying whether $N_1$ is indeed the dark matter.

\section*{Acknowledgements}

MRL would like to thank Jes\'us Zavala, Tamar Meshveliani, Esra Bulbul, Michael Boylan-Kolchin, Francis-Yan Cyr-Racine, Simona Vegetti and the anonymous referee for useful comments.

\section*{Data Availability}

The original codes for \citetalias{Venumadhav16} and \citetalias{Ghiglieri15} are available at \href{https://github.com/ntveem/sterile-dm}{https://github.com/ntveem/sterile-dm} and \href{http://www.laine.itp.unibe.ch/dmpheno/}{http://www.laine.itp.unibe.ch/dmpheno/} respectively. 



\bibliographystyle{mnras}



\appendix

\section{The effective number of relativistic degrees of freedom}
\label{app:neff}

\begin{figure}
    \centering
    \includegraphics[scale=0.33]{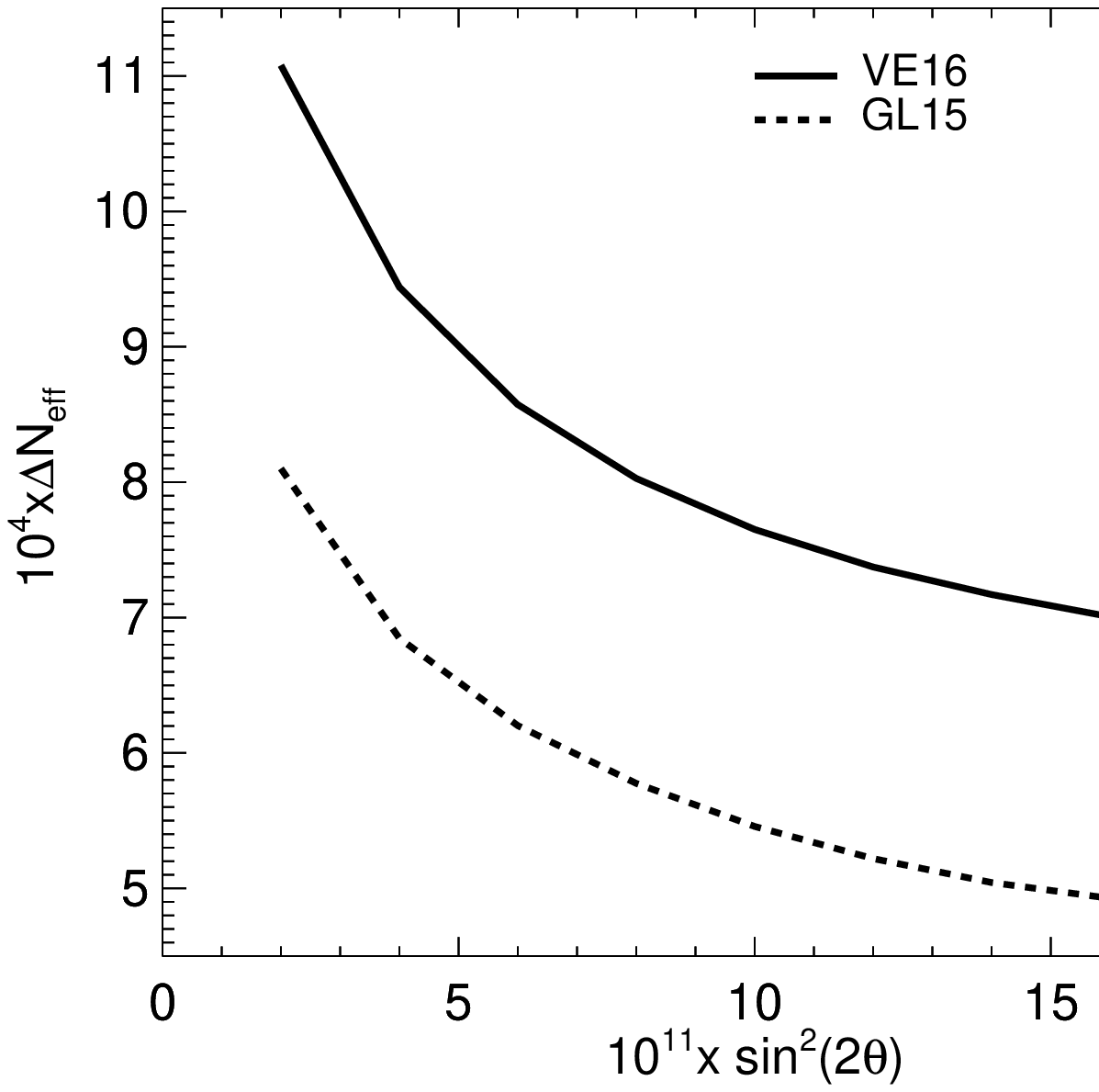}
    \caption{The relationship between \stt and $\Delta N_\rmn{eff}$ for the two models when the {\sc class} code is applied to calculate the linear matter power spectrum. The \citetalias{Venumadhav16} result is shown as a solid line and the \citetalias{Ghiglieri15} result is shown as a dashed line.}
    \label{fig:s2neff}
\end{figure}

An interesting corollary of the late transition from relativistic to non-relativistic properties for $N_{1}$ is the impact on $N_\rmn{eff}$, and whether this value is high enough to be measurable with current or future cosmological observations. We therefore compute values of the excess degrees of freedom due to $N_1$, labelled $\Delta N_\rmn{eff}$ for the \citetalias{Venumadhav16} and \citetalias{Ghiglieri15} models as a function of \stt; we apply the {\sc class} code for this task. We present the results in Fig.~\ref{fig:s2neff}.

The value of $\Delta N_\rmn{eff}$ is inversely correlated with \stt, in line with the initial $N_1$ temperature distribution and the subsequent matter power spectrum cutoff. The highest value of $\Delta N_\rmn{eff}$ for \citetalias{Venumadhav16} is $1.1\times10^{-3}$, at \stt$=2\times10^{-11}$ and smoothly drops to $6.8\times10^{-4}$ at \stt$=2\times10^{-10}$. The same relation shape occurs for \citetalias{Ghiglieri15}, albeit with a lower normalization such that the highest value is $8.1\times10^{-4}$. These results are lower than the $\pm0.17$ uncertainty in $N_\rmn{eff}$ derived in \citet{PlanckCP20}, and therefore it is unlikely than any experiment can probe this result in the near future.  


\bsp	
\label{lastpage}
\end{document}